\documentclass[10pt,twocolumn,letterpaper]{IEEEtran}

\usepackage{amssymb,epsf,epsfig,graphicx,latexsym,amsmath}

\usepackage{graphicx}
\usepackage{amssymb}
\usepackage{times}

\title{Performance of  Ultra-Wideband Impulse Radio \\
  in Presence of Impulsive Interference}

\author{Bo\v zidar Radunovi\'c$^\star$\thanks{$^\star$ Microsoft
Reseach, 7 JJ Thomson Avenue, Cambridge, CB3 0FB,
UK. bozidar@microsoft.com}
\hspace{1em}
Jean-Yves Le Boudec$^\ddagger$\thanks{$\ddagger$
School of Computer and Communication Sciences, EPFL,
CH-1015 Lausanne, Switzerland. jean-yves.leboudec@epfl.ch}
\hspace{1em}
Raymond Knopp$\bigtriangleup$\thanks{$\bigtriangleup$, Mobile Communications Laboratory, Institut Eur\'ecom, 06904 Sophia Antipolis, France. knopp@eurecom.fr}
}

\begin{document}


\newtheorem{lemma}{Lemma}
\newtheorem{theorem}{Theorem}
\newtheorem{definition}{Definition}
\newtheorem{corollary}{Corollary}

\newtheorem{bbb}{Corollary}

\newtheorem{claim}{Claim}
\newtheorem{proposition}{Proposition}
\newtheorem{heuristic}{Heuristic}

\newcommand\FN{{\mathcal{N}}}
\newcommand\FT{{\mathcal{T}}}
\newcommand\FF{{\mathcal{F}}}
\newcommand\FX{{\mathcal{X}}}
\newcommand\FY{{\mathcal{Y}}}
\newcommand\FP{{\mathcal{P}}}
\newcommand\FS{{\mathcal{S}}}
\newcommand\FXa{\bar{\mathcal{X}}}
\newcommand\FC{{\mathcal{C}}}

\newcommand\sN{{\{1,\cdots,N\}}}
\newcommand\sL{{\{1,\cdots,L\}}}
\newcommand\sS{{\{1,\cdots,S\}}}

\newcommand\bp{{\bf{p}}}
\newcommand\bq{{\bf{q}}}
\newcommand\br{{\bf{r}}}
\newcommand\bs{{\bf{s}}}
\newcommand\bP{{\bf{P}}}
\newcommand\bQ{{\bf{Q}}}
\newcommand\bx{{\bf{x}}}
\newcommand\by{{\bf{y}}}
\newcommand\bz{{\bf{z}}}
\newcommand\bbf{{\bf{f}}}
\newcommand\bg{{\bf{g}}}
\newcommand\bc{{\bf{c}}}
\newcommand\bt{{\bf{t}}}
\newcommand\bR{{\bf{R}}}
\newcommand\bH{{\bf{H}}}
\newcommand\bY{{\bf{Y}}}
\newcommand\bG{{\bf{G}}}
\newcommand\bU{{\bf{U}}}
\newcommand\bV{{\bf{V}}}
\newcommand\bW{{\bf{W}}}
\newcommand\bI{{\bf{I}}}
\newcommand\balpha{{\mbox{\boldmath$\alpha$}}}
\newcommand\blambda{{\mbox{\boldmath$\lambda$}}}
\newcommand\bomega{{\mbox{\boldmath$\omega$}}}
\newcommand\bOmega{{\mbox{\boldmath$\Omega$}}}
\newcommand\bmu{{\mbox{\boldmath$\mu$}}}
\newcommand\bbeta{{\mbox{\boldmath$\eta$}}}
\newcommand\brho{{\mbox{\boldmath$\rho$}}}
\newcommand\be{{\bf{e}}}
\newcommand\bu{{\bf{u}}}
\newcommand\bv{{\bf{v}}}
\newcommand\bw{{\bf{w}}}
\newcommand\bh{{\bf{h}}}
\newcommand\bb{{\bf{b}}}
\newcommand\bd{{\bf{d}}}

\newcommand\hh{{\hat{h}}}
\newcommand\hC{{\hat{C}}}
\newcommand\hD{{\hat{D}}}

\newcommand\bhc{{\hat{\bf c}}}
\newcommand\bhh{{\hat{\bf h}}}
\newcommand\bhf{{\hat{\bf f}}}

\def\Reals{\mathbb{R}}
\def\len{\mathrm{len}}

\newcommand\bxa{{\bar{\bf{x}}}}
\newcommand\xa{{\bar{x}}}
\newcommand\bpa{{\bar{\bf{p}}}}
\newcommand\pa{{\bar{p}}}
\newcommand\Ra{{\bar{R}}}

\newcommand\src{\mathrm{Src}}
\newcommand\dst{\mathrm{Dst}}
\newcommand\err{\mathrm{err}}

\newcommand\sor{{\; \vee\; }}
\newcommand\sand{{\; \wedge\; }}

\newcommand{\mor} {\mathrm{\;  or \; }}
\newcommand{\mand} {\mathrm{\;  and \; }}

\newcommand{\sinc} {\,\mathrm{sinc}\,}

\newcommand{\comb}[2]{\left(\begin{array}{c}#1\\#2\end{array}\right)}

\def\eps{\epsilon}
\def\RR{\mathbb{R}}
\def\NN{\mathbb{N}}
\def\EE{\mathbb{E}}
\def\PP{\mathbb{P}}


\newcommand\bsection{\section}
\newcommand\bsubsection {\subsection}
\newcommand\bsubsubsection {\subsubsection}

\newcommand{\mycaption}[1]{\caption{\small \sffamily #1}}


\newcommand{\fref}[1]{Figure~\ref{#1}}
\newcommand{\sref}[1]{Section~\ref{#1}}
\newcommand{\cref}[1]{Chapter~\ref{#1}}
\newcommand{\coref}[1]{Corollary~\ref{#1}}
\newcommand{\eref}[1]{Equation~(\ref{#1})}
\newcommand{\tref}[1]{Table~\ref{#1}}
\newcommand{\dref}[1]{Definition~\ref{#1}}
\newcommand{\lref}[1]{Lemma~\ref{#1}}
\newcommand{\thref}[1]{Theorem~\ref{#1}}
\newcommand{\pref}[1]{Proposition~\ref{#1}}
\newcommand{\pgref}[1]{Page~\pageref{#1}}
\newcommand{\qref}[1]{Question~\ref{#1}}

\newcommand{\pfref}[1]{Figure~\ref{#1} on page~\pageref{#1}}
\newcommand{\psref}[1]{Section~\ref{#1} on page~\pageref{#1}}
\newcommand{\pcref}[1]{Chapter~\ref{#1} on page~\pageref{#1}}
\newcommand{\pcoref}[1]{Corollary~\ref{#1} on page~\pageref{#1}}
\newcommand{\peref}[1]{Equation~(\ref{#1}) on page~\pageref{#1}}
\newcommand{\ptref}[1]{Table~\ref{#1} on page~\pageref{#1}}
\newcommand{\pdref}[1]{Definition~\ref{#1} on page~\pageref{#1}}
\newcommand{\plref}[1]{Lemma~\ref{#1} on page~\pageref{#1}}
\newcommand{\pthref}[1]{Theorem~\ref{#1} on page~\pageref{#1}}
\newcommand{\ppref}[1]{Proposition~\ref{#1} on page~\pageref{#1}}
\newcommand{\pqref}[1]{Question~\ref{#1} on page~\pageref{#1}}
\newcommand{\exref}[1]{Example~\ref{#1} on page~\pageref{#1}}


\newcommand\bprcv{{{\bf p}_{\mbox{rcv}}}}
\newcommand\prcv{{p_{\mbox{rcv}}}}
\newcommand\allp{{\{\bp^n(s)\}_{1,\cdots,N}}}
\newcommand\allps{{\{\bp^{n*}(s)\}_{1,\cdots,N}}}
\newcommand\sNs{{\{1,\cdots,N(s)\}}}

\newcommand\PM{P^{MAX}}
\newcommand\bPM{{\bf P}^{MAX}}
\newcommand\PaM{\overline{P}^{MAX}}
\newcommand\bPaM{\overline{\bf P}^{MAX}}
\newcommand\FaM{\overline{F}^{MIN}}
\newcommand\bFaM{\overline{\bf F}^{MIN}}
\newcommand\bXaM{\overline{\bf X}^{MIN}}
\newcommand\bSNR{\mathbf{SINR}}
\newcommand\bSINR{\mathbf{SINR}}
\newcommand\SNR{\mathrm{SINR}}
\newcommand\FPa{\overline{\mathcal P}}

\newcommand\nh{{r}}

\newcommand\interf{{Z}}
\newcommand\wn{{\eta}}

\newcommand\vecr{{\bf r}}
\newcommand\vech{{\bf h}}
\newcommand\vecz{{\bf z}}

\renewcommand{\mycaption}[1]{\caption{\scriptsize \sffamily #1}}

\renewcommand\bsection {\section}
\renewcommand\bsubsection {\subsection}

\maketitle

\begin{abstract}
We analyze the performance of coherent impulsive-radio (IR)
ultra-wideband (UWB) channel in presence of the interference
generated by concurrent transmissions of the systems with the same
impulsive radio. We derive a novel algorithm, using Monte-Carlo
method, to calculate a lower bound on the rate that can be achieved
using maximum-likelihood estimator. Using this bound we show that
such a channel is very robust to interference, in contrast to the
nearest-neighbor detector.

\end{abstract}

\section{Introduction}

In this work we consider an impulse-radio ultra-wideband (IR-UWB)
system. This type of communication is being actively developed
because of its promising features. It generates low-power
transmissions over a large bandwidth, hence achieves high bit rates
with low power consumption. Other advantages of IR-UWB are low cost,
multi-path immunity and precise ranging capabilities. The IR-UWB
physical layer is adopted as a standard for 802.15.4a personal-area
networks (PAN) in the 3-5GHz band.

The current IR-UWB receivers are designed to work with Gaussian
white noise, and they use nearest-neighbor decoding which, in case
of Gaussian white noise, corresponds to maximum-likelihood decoding
(MLE). However, in a networking environment, where there are several
concurrent transmissions of IR-UWB devices, the interference is very
impulsive and nearest-neighbor decoding is suboptimal. Interference
may occur because of several competing piconets, or during a random
access phase in the same piconet.

We are interested in calculating the achievable rates of a coherent
IR-UWB channel in presence of white noise and impulsive interference
generated by a network of IR-UWB transmitters. Instead of
nearest-neighbor, we focus on the optimal, MLE decoder. Although we
cannot calculate the maximum achievable rate explicitly, we give a
novel lower bounds on the achievable rate. We also present a simple
upper bound and we use these bounds to evaluate the performance of
the channel.

We further consider an example of a channel with a single impulsive
interferer. We show that if one uses an MLE decoder, in many cases it
can mitigate the effect of impulsive interference, especially when
the interference is strong. This is in contrast to IR-UWB receivers
with nearest-neighbor decoding which are highly penalized by strong
impulsive interference \cite{Durisi03}.

In \sref{sec:relwork} we present related work. In
\sref{sec:assumptions} define system assumptions. We derive the
upper and lower bounds in \sref{sec:bounds}, we illustrate their
performance in \sref{sec:results} and we conclude.

\section{Related Work}
\label{sec:relwork}

The performance of a channel with additive, non-Gaussian noise and
nearest neighbor decoder is discussed in \cite{lapidoth96}. In
particular, the case with impulsive interference is discussed in
\cite{Durisi02, Durisi03} where the authors show that the Gaussian
approximation of the interference is not correct and give a
numerical model to evaluate the bit-error rate in homogeneous
settings. A similar problem is found in \cite{Merz05}, where fast
and more efficient methods are developed to handle heterogeneous
cases and with multipath channels. However, in all of these works,
the author consider the nearest neighbor decoder, which is not the
optimal one when the interference is not Gaussian.

Our work is inspired by \cite{Souilmi03I}, which calculates an upper
and a lower bound on achievable rates of a non-coherent IR-UWB
channel. This work is extended in \cite{Souilmi03D} to evaluate the
performance of different non-optimal detectors. A similar analysis
is done in \cite{Luo06} for transmitted-reference IR-UWB radio.
Nevertheless, in all of these works the authors do not consider the
effects of impulsive interference from concurrent transmission of
the same type of radios.  An appoximate implementation of MLE
decoder for multi-user interference channel is given in
\cite{Steiner05}. However, to the best of our knowledge, we are the
first to numerically calculate bounds on the achievable rate of such
a channel.

\section{System Assumptions}
\label{sec:assumptions}

\subsection{Channel model}

We consider a set of $I+1$ nodes. Node 1 communicates with node 0
while the other $I-1$ nodes generate impulsive interference from
concurrent transmissions. All nodes use IR-UWB for communication,
meaning that they transmit a sequence of very short pulses. All signals and channels are assumed to be real-valued. The
received signal at node 0 is equal to
\begin{eqnarray}
  r(t) &=& \sum_{n=-\infty}^\infty u_1[n] A_{1} h_{1}(t - n T_s) + z(t) \nonumber \\
  &+& \sum_{i = 2}^{I+1} \sum_{n=-\infty}^\infty u_i[n]
  A_{i} h_{i}(t - n T_s + \Delta_{in}), \label{eq:time_domain}
\end{eqnarray}
where $u_{in}$ is the $n$-th transmitted symbol by node $i$, $A_{i}$
is the received amplitude of a pulse at node 0, transmitted from node
$i$, $\Delta_{in}$ is the time shift (due to asynchronicity) between
interferer $i$ and destination 0 for the $n$-th symbol, $T_s$ is the symbol duration, and $h_{i}$
is the normalized channel impulse response of a channel from node
$i$ to node 0, which is described next.

When a transmitted signal propagates from one point to another, it
travels over multiple paths. If a perfect impulse $\delta(t)$ is
transmitted, the received signal will be
\[
   h^p(t) = \sum_{s=1}^L a_s \delta(t - \tau_s),
\]
where $a_s$ is the attenuation and $\tau_s$ is the delay of $s$-th
path ($s$ taking values from 1 to $L$). We assume that $h_i(t)$ is normalized so that$\sum_{s=1}^L a_s^2= 1$.

The received impulse response $h^p(t)$ is further filtered to the
system bandwidth $W$, and the received signal is
\begin{eqnarray*}
  h(t) &=& \sqrt{2W}\sum_{s=1}^L a_s \sinc 2W(t - \tau_s).
\end{eqnarray*}

We further assume that we sample the channel impulse response at
$2W$, and we have the following samples
\[
   h_m = h\left({m\over{2 W}}\right) = \sqrt{2W}\sum_{s=1}^L
         a_s \sinc 2W({m\over 2 W} - \tau_s),
\]
for $m=1,\cdots,M$ where $M$ is the number of samples. Since the
number of paths is very large, $h_m$ is a sum of a large number of
random variables thus it can be assumed Gaussian. However, different
samples are not independent. Therefore we assume that each $\bh_i =
(h_{i1},\cdots,h_{iM})$ is a multi-variate Gaussians and
$\{\bh_i\}_i$ are i.i.d. with zero mean and covariance matrix $T_i$.

Channels $\bh_i$ are independent but their statistics depend on the
time jitter $\Delta_{in} = 0$. In an unlikely case that interferer
$i$ is symbol-level synchronized to a receiver (that is $\Delta_{in}
= 0$), the receiver will receive the full energy of this interferer.
Then, the distribution of $\bh_i$ will be the same as the
distribution of $\bh_1$, for the intended transmitter, that is,
perfectly synchronized as well. Symbol-level synchronized interferer
is thus a worst case approximation. Although we can readily use our
model for arbitrary $\Delta_{in}$, we are interested in deriving a
lower-bound, hence we will assume $\Delta_{in} = 0$ for all $i,n$.
Hence, we will assume that all channel responses $\bh_i$ have the
same covariance matrix $T_i = T$.

For each transmitted symbol we have $M$ channel samples. Each sample
can be interpreted as one dimension of a channel. We can thus
formulate the corresponding channel as discrete vector channel, and we
have the following channel model
\begin{equation}
   \br[n] = u_{1}[n] A_{1}[n] \bh_{1}[n]
          + \sum_{i=2}^{I+1} u_i[n] A_i[n] \bh_i[n]
          + \bz[n]. \label{eq:vecchan_simple}
\end{equation}
where $\bh_i[n]$ are the samples of channel impulse responses during
transmission of symbol $n$ and $\bz[n]$ are samples of white noise
hence i.i.d. Gaussian.

The typical channel coherence time for a UWB channel is of order of
tenths of milliseconds \cite{Tse05}. In practice, this means that
the channel is constant throughout the duration of a packet and, in
\eref{eq:vecchan_simple}, $\bh_i[n] = \bh_i$ are constant in $n$.
Furthermore, since we consider coherent communication, we assume the
receiver knows transmitter's channel $\bh_1$ but it does not know
the interferer s' channels $\{\bh_i\}_{i\not=1}$.

\subsection{Transmitter and Receiver Structure}

It has been shown that an efficient IR-UWB system needs to transmit
infrequent pulses \cite{Telatar00} due to a very low available
average transmission power. This in turn facilitates multi-user
communication. One implementation of such a principle are for
example time-hopping (TH) codes \cite{Win00}.

We use a generic signalling model and assume that $\{u_i[n]\}_n$ are
i.i.d. random variables with $P(u_i[n] = 1) = \eta_i$ and $P(u_i[n]
= 0) = 1-\eta_i$.  Variable $\eta_i$ thus denotes the duty-cycle of
node $i$. This model is more general than the TH model as it does
not impose any dependency between symbols, which is needed in case
of TH due to implementational constraints.

We also assumed that the transmitted symbol can be only $0$ or
$1$. This is not the most general type of modulation. For example,
having $u_i[n]\in\{-1,0,1\}$ would allow us to transmit additional
information in the symbol phase.
However, for the simplicity of the presentation we assume
$u_i[n]\in\{0,1\}$ and we note that the result can easily be extended
to different modulation schemes.

The average power during transmission is upper-bounded by
$P_i^{MAX}$, where the value of $P_i^{MAX}$ is specified by
regulations. Since the probability of transmitting 1 is $\eta_i$,
the amplitude of a pulse is bounded by $A_i^2 \leq P_i^{MAX} /
\eta_i$.

We shall not assume any particular coding or detection techniques
for our system. We will use random coding to derive a lower bound
and Shannon capacity to derive an upper bound on the achievable
rates. The precise descriptions are given next.

\section{Bounds on Achievable Rate}
\label{sec:bounds}

We want to derive bound on the achievable link's physical data rate,
given the received signal power and the received powers of
interferers. For that matter, we will consider the discrete vector
channel model described in (\ref{eq:vecchan_simple}). We will first
present a simple upper-bound, and then we will derive a novel
lower-bound, which is the main result of our paper.

\subsection{Upper-bound}

To derive an upper-bound we consider the information-theoretic
capacity of the channel, constrained to the binary input alphabet. We can represent our channel as an $M$-dimensional vector channel
\begin{equation}
   \bR = U_{1} A_{1} \bh_{1} + \sum^I_{i>1}  U_i A_i \bh_i + \bz,
\end{equation}
where $\{U_i\}_{i=1\cdots I}$ are i.i.d Bernouli random variables with $P(U_i = 1) =
\eta_i$, $\bz = \{z_m\}_{m=1\cdots M}$ are i.i.d Gaussian with
variance $\sigma_W^2$, $\{A_i\}_{i=1\cdots I},\{\bh_{im}\}_{i=1\cdots
  I, m=1\cdots M}$ are known signal attenuations and unknown (except
for $\bh_1$) but constant channel fadings from transmitter $i$ to the
destination respectively.

We upper-bound the capacity of the channel assuming the receiver knows
the received symbols $\{\bu_i\}_{i>1}$ from the interferers. Since it
can then perfectly estimate the interference and extract it, the only
remaining noise is the white noise $\bz$.

We use the notation $\bH = \{\bh_i\}_{i=1,\cdots,I}$ and $\bU_{-1} =
\{U_i\}_{i=2,\cdots,I}$. We can write the upper-bound as
\begin{eqnarray}
  C_u &=& I(u_1; \bR \,|\, \bh_1)\\
    &\leq& I(u_1; \bR \,|\, \bU_{-1}, \bH) \\
    &=& \EE_{\bh_1,U_1,\bz} \left[-\log\left(\sum_{v=\{0,1\}} P(U_1=v)
        e^E\right)\right]\\
  E &=& -\sum_{m=1}^M\frac{(z_m-(v-U_1)A_1h_{1m})^2-z_m^2}{2\sigma_W^2}
\end{eqnarray}
The result is similar to the result from \cite{Souilmi03D} for a
non-coherent channel. This bound can be easily calculated using
e.g. Monte-Carlo simulations.

\subsection{Lower-bound}

\subsubsection {Threshold Decoding}

Next, we will derive a lower-bound on achievable rates using a
practical decoding scheme. We suppose the source sends data in packets
of length $N$, where $N$ is assume large. Each packet has a coding
rate $C_R$ associated with it, yielding error probability of decoding
$P(\err)$. We consider an upper bound on $P(\err)$ using random coding
bound technique.

Suppose packets are coded using a random codebook $\FC \subseteq
\{0,1\}^{N}, ||\FC|| = 2^{N C_R}$. A source, knowing the channel state
$\bh_1$, sends codeword $\bu_1$ and a destination receives $\bR =
\{\br[n]\}_{n=1,\cdots,N}$, as described in
(\ref{eq:vecchan_simple}). The optimal decoder is the maximum
likelihood (MLE) decoder which selects the codeword $\bu_1$ that
maximizes the likelihood $P(\bR\,|\,\bu_1,\bh_1)$. However, the
performance of the maximum likelihood decoder is hard to
analyze. Since we are interested in an upper-bound on the probability
of error, we shall consider a simple threshold decoding scheme, based
on an arbitrary threshold $\theta$. If the likelihood
$P(\bR\,|\,\bu_1,\bh_1) > \theta$ for only $\bu_1\in\FC$, then the
decoding is successful.  Otherwise, it fails.

\subsubsection {Performance Analysis}

We start by giving all the notation we will be using: $\bH =
\{\bh_i\}_{i=1,\cdots,I}$, $\bH_{-1} = \{\bh_i\}_{i=2,\cdots,I}$,
$\bR = \{\br[n]\}_{n=1,\cdots,N}$, $\bu_i =
\{u_i[n]\}_{n=1,\cdots,N}$, $\bU = \{\bu_i\}_{i=1,\cdots,I}$,
$\bU_{-1} = \{\bu_i\}_{i=2,\cdots,I}$ and $\bU[n] =
\{u_i[n]\}_{i=1,\cdots,I}$. We will also use a short notation for
$P(\bY\,|\,\bv,\bh) = P(\bR = \bY\,|\,\bu_1 = \bv,\bh_1 = \bh)$ and
$P(\bY\,|\,\bV,\bh) = P(\bR = \bY\,|\,\bU = \bV,\bh_1 = \bh)$.

We first need to choose the threshold $\theta$ which will yield good
performance. Ideally, $\theta(\bv, \bh_1)$ is a function of the
(unknown) transmitted codeword $\bv$ and a channel-state $\bh_1$, and
we shall choose it to minimize the probability of false-negative
$P(P(\bR\,|\,\bv, \bh_1) < \theta(\bv, \bh_1) \,|\, \bu_1=\bv,
\bh_1)$. We will show later that the optimal $\theta$ does not depend
on the choice of $\bv, \bh_1$.

The noise and the interferences are ergodic processes hence for a
large packet size $N$ we have that $P(P(\bR\,|\,\bv, \bh_1) >
\theta(\bv, \bh_1)| \bu_1 = \bv, \bh_1) \to 1$ if
\begin{eqnarray}
  \theta(\bv,\bh_1) &=& (1-\eps) \EE_{\bR}(P(\bR\,|\,\bv, \bh_1) \,|\, \bu_1=\bv, \bh_1)\\
    &=& (1-\eps) \int_\bY P(\bR = \bY\,|\,\bv, \bh_1)^2 d\bY \label{eq:theta}
\end{eqnarray}
for any $\eps>0$. We will choose $\theta(\bv,\bh_1) = \int_\bY
P(\bR=\bY\,|\,\bv, \bh_1)^2 d\bY$ (i.e. $\eps = 0$), and assume
further $P(P(\bR\,|\,\bv, \bh_1) > \theta(\bv, \bh_1)| \bu_1=\bv,
\bh_1) = 1$. We next show that $\theta(\bv, \bh_1)$ does not
actually depend on $\bv, \bh_1$, hence we can write $\theta(\bv,
\bh_1) = \theta$. {\proposition \label{prop:sym} The following
integral
$$
   p(||\bv-\bw||, \bh_1) = \int_\bY P(\bR=\bY\,|\,\bv, \bh_1)
      P(\bR=\bY\,|\,\bw, \bh_1) d\bY
$$
depends only on $||\bv-\bw||$. Also, $\theta(\bv, \bh_1)$ depends neither
on $\bv$ nor on $\bh_1$.}

\begin{proof}
Let us denote with $\bQ[n] = \sum_{i=2}^I u_i[n] A_i[n]
\bh_i[n] + \bz[n]$. Then, $P(\bR=\bY\,|\,\bv, \bh_1) =
P(\bigcup_{n=1\cdots N} \bQ[n] = \bY[n] - \bv[n] A_1 \bh_1)$ and
\begin{eqnarray*}
  \lefteqn{\int_\bY P(\bR=\bY\,|\,\bv, \bh_1) P(\bR=\bY\,|\,\bw, \bh_1) d\bY =}\\
    && \int_\by P\left(\bigcup_{n=1\cdots N}
       \bQ[n] = \bY[n]\right) \times \\
    &\times& P\left(\bigcup_{n=1\cdots N} \bQ[n] = \bY[n] -
       (\bw[n]-\bv[n]) A_1 \bh_1\right) d\by.
\end{eqnarray*}
The distribution of the vector $\{\bQ[n]\}_{n=1\cdots N}$ is by
definition symmetric and invariant to a permutation of its elements,
hence the value of the integral depends only on $||\bv-\bw||$.
Furthermore, if $||\bv-\bw|| = 0$, as in (\ref{eq:theta}), then the
integral does not depend on $\bh_1$ either.
\end{proof}

Now we are interested in the probability of error of decoding a
random transmitted codeword. We consider a random codebook $\FC$,
and from there select a random codeword $\bv$ to transmit. Note that
$P(\bv=\bomega \,|\, \FC=C) = 2^{-C_R N} 1\{\bomega\in C\}$ since
all the codewords from $\FC$ are equiprobable. The probability of
error can be bounded by the union bound as
\begin{eqnarray}
  \lefteqn{P(\err | \bh_1) \leq}\\
  &\leq& \EE_{\FC,\bv\in\FC} \left[ \sum_{\bomega\in\FC, \bomega \not= \bv}
        P(P(\bR\,|\,\bomega, \bh_1) > \theta \,|\, \bv, \bh_1) \right] \\
    &=& 2^{C_R N} \EE_{\bv, \bw, \bv\not=\bw}\left[P(P(\bR\,|\,\bw, \bh_1) >
      \theta \,|\, \bv, \bh_1)\right] \label{eq:tresdec}
\end{eqnarray}
where $\bv,\bw$ are two randomly choosen codewords from a random codebook.

Next, using Markov inequality, we bound
\begin{eqnarray}
  \lefteqn{P(P(\bR\,|\,\bw, \bh_1) > \theta \,|\, \bu_1=\bv, \bh_1) \leq}\\
  &\leq& \frac{1}{\theta}
     \EE_\bR\left[ P(\bR\,|\,\bw, \bh_1) \,|\, \bv, \bh_1 \right]\\
  &=& \frac{p(||\bv-\bw||, \bh_1)}{\theta} \label{eq:markov}
\end{eqnarray}
where the last equation follows from \pref{prop:sym}. The Markov
bound is the best bound we can use knowing only the mean of a random
variable, and numerical results in \sref{sec:results} show that the
bound is useful for performance evaluation of the channel.

Random codewords $\bv,\bw$ can be assumed independent in a large
codebook (when $C_R N$ is large). We then have that $P(||\bv-\bw|| =
d) = \comb{N}{d} (2\eta_1(1-\eta_1))^d (\eta_1^2 +
(1-\eta_1)^2)^{N-d}$.

Next, let $e_d$ be a vector with $d$ ones and $n-d$ zeros and let us
denote $J(\bV,\bW,\bh_1) = \int_\bY P(\bR=\bY\,|\,\bV,\bh_1)
P(\bR=\bY\,|\,\bW,\bh_1) d\bY$. Then from (\ref{eq:tresdec}) and
(\ref{eq:markov}) we have
\begin{eqnarray}
  p(d) &=& \EE[ J([\be_d,\bU_{-1}],[\be_0,\bV_{-1}],\bh_1)], \label{eq:pd}\\
  P(\err) &\leq& \frac{2^{C_R N}}{\theta}
    \sum_{d=0}^N P(||\bu-\bv|| = d) p(d) \label{eq:pebound}
\end{eqnarray}
where $\bU_{-1} = \{u_i\}_{i=2,\cdots,I}, \bV_{-1} =
\{v_i\}_{i=2,\cdots,I}$ are random codewords transmitted by
interferers. We can express $J(\bU,\bV,\bh_1)$ in a closed-form, as
explained in Appendix, and calculate the mean using Monte-Carlo
simulations. Since $p(0,\bh_1) = \theta$, we can use the same
procedure to calculate $\theta$ (note that in addition $\theta$ does
not depend on $\bh_1$). Details on Monte-Carlo simulations are given
in \sref{sec:MC}.

From (\ref{eq:pebound}) we can obtain a lower-bound $C_l$ on the
communication rate $C_R$. When $N\to\infty$ we can obtain
arbitrarily small $P(\err)$ using communication rate
\begin{equation}
  C_l = -\frac{1}{N} \log_2\left(\sum_{d=0}^N P(||\bu-\bv|| = d)
      \frac{p(d)}{\theta}\right). \label{eq:lbound}
\end{equation}

\section{Numerical Results}
\label{sec:results}

In this section we first discuss the reliability of the Monte-Carlo
simulation, and then illustrate the results on a channel with a single
interferer.

\subsection{Monte-Carlo Simulations}
\label{sec:MC}

We calculate $\theta$ and $\sum_{d=0}^N P(||\bu-\bv|| = d) p(d)$ using
Monte-Carlo simulations, averaging over many random samples of
$\bV_{-1},\bU_{-1}$ and $\bh_1$.

In the case of $\theta$, we verify that the samples obtained by
Monte-Carlo fit the Gaussian distribution well. This allows us to
calculate confidence intervals of the simulation \cite{LeBoudec06},
and in all cases the relative confidence intervals are smaller than
10\%.

In case of $\sum_{d=0}^N P(||\bu-\bv|| = d) p(d)$ the samples are no
longer Gaussian, but we verify that a log transform is Gaussian. There
is a simple intuitive explanation for this. For very small $d$
($d=1,2$) the candidate and the transmitted codewords are similar, the
probability of error (estimated through $p(d)$) is high. However,
there are a few such codewords. On the contrary, for large $d$, $p(d)$
is small, but there are a lot of such words.

In all the simulations we find that the relative confidence
intervals for the error probability of decoding $P(\err)$ are
smaller than 50\%. We are interested in $C_l$ (\ref{eq:lbound})
which is of order of $\log_2(P(\err))$. Since the values of interest
of $\log_2(P(\err))$ are smaller than -20, we can see that the
relative confidence for $C_l$ is approximately 5\%.

We also find that the lower-bound coincides with the upper-bound in
the case of purely Gaussian interference (since our lower bound then
coincides with a well-known random-coding bound for AWGN channels).

\subsection{Single Interferer Channel}
\label{sec:schannel}
\begin{figure}[!tb]
     \includegraphics[width=0.8\linewidth]{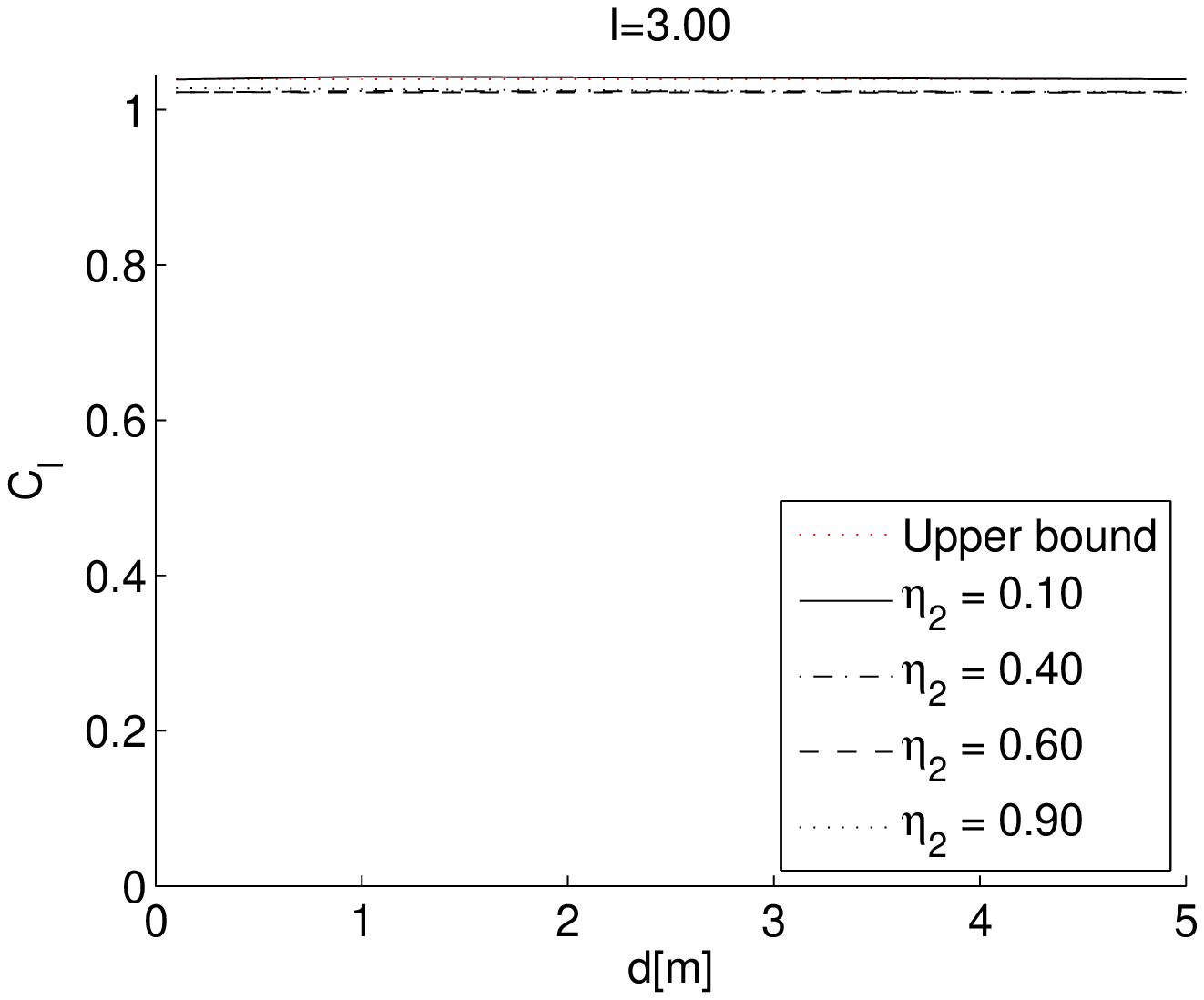}
     \includegraphics[width=0.8\linewidth]{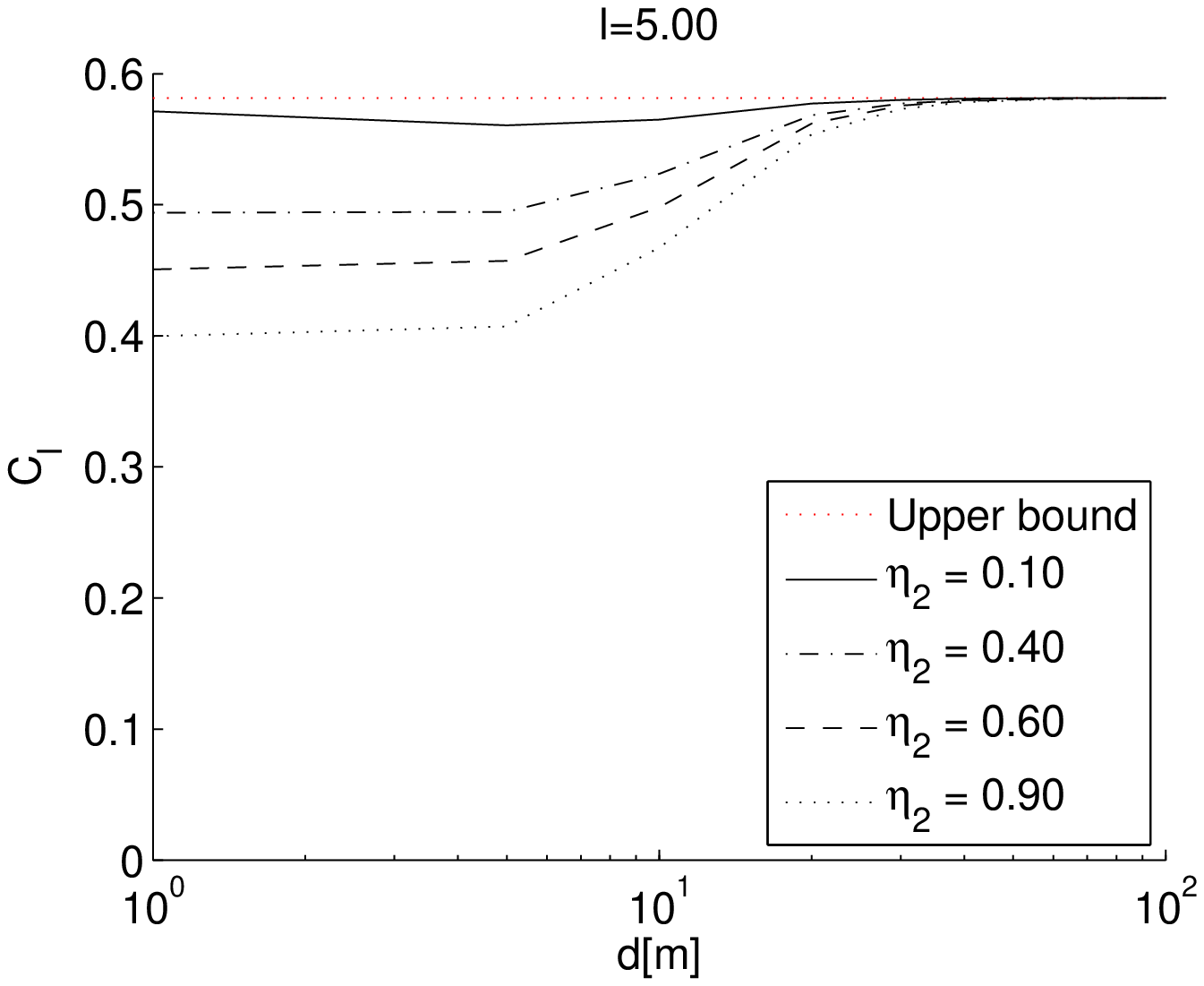}
     \includegraphics[width=0.8\linewidth]{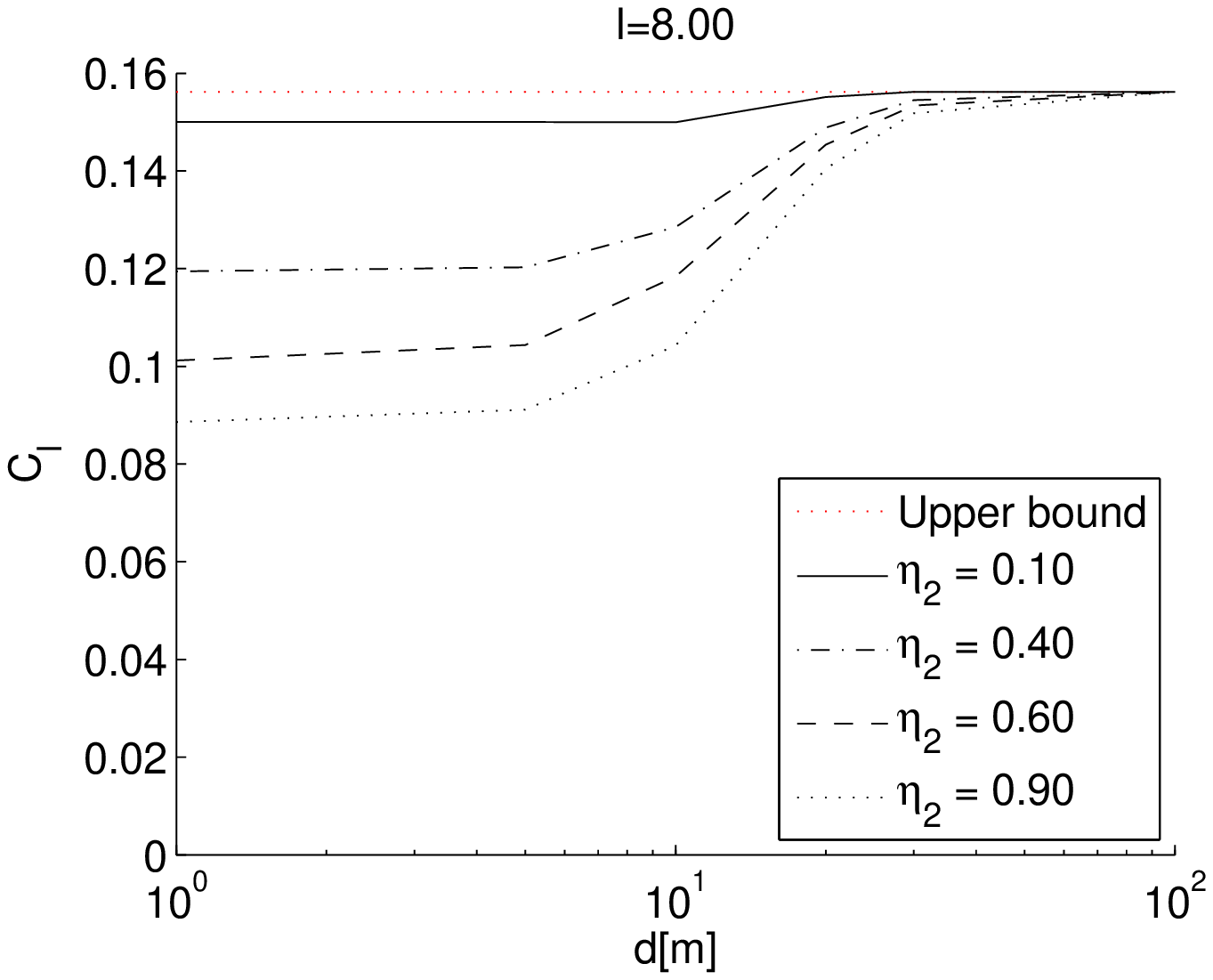}
\caption{\label{fig:eta_fix} Upper and lower bounds on achievable
rates of channels with a single interferer for $l=3$m, $l=5$m, and
$l=8$m. The lower bound is given for different values of $\eta_2$
(the upper bound is independent of $\eta_2$). The upper bound
coincides with the lower-bound for $d=100$m.}
\end{figure}

In order to illustrate our results, we consider a channel with a
single interferer ($I=2$).  We take $N=80$ which is around the maximum
packet size we can use to have reasonable simulation times. We also
take $M=5$ corresponding to a 5-tap receiver, which is also aroung the
maximum $M$ we can simulate.

For channel statistics we use the measurements from \cite{Souilmi03D}
which says that the tap energy drops linearly with the tap delay.
Approximately 14\% of the total energy is in the first 5 taps. We also
assume here that taps are independent.

In this section we are interested in comparing performance of channels
for different channel parameters. In order to avoid additional
unnecessary variance in our results, we will assume here that $\bh_1$
is given. Similar results hold for different values of $\bh_1$ (hence
will hold for the average channel realisations as well).

We further suppose that the distance between receiver 1 and receiver
0 is $l$ and the distance between interferer 2 and receiver 1 is
$d$. Both 1 and 2 transmit with transmitting power $P_{\mbox{trans}}
= 0.1$mW. The average received power at distance $l$ is
$P_{\mbox{rcv}}(l) = P_{\mbox{trans}} \, b \, l^{-\alpha}$ where $b
= 10^{-5.5}, \alpha = 3.3$ are taken from \cite{ghassemzadeh03}.
White-noise power is $\sigma_W^2 = 10^{-13}W$.  The maximum
communication range of such a system is around 10m, so in our
simulations we use link sizes of $l=$3,5,8m.

In order to illustrate the result we consider a scenario in which we
fix $\eta_1 = 0.5$ and for different $l$ we vary $d$ and $\eta_2$.
The results are shown in \fref{fig:eta_fix}. We see that for $l=3$m
the achievable rate is insensitive to the interference. For $l=5$m
and $l=8$m the rate decreases when the interferer is closer than $d
< 30$m. However, it drops until $d=5$m, and for smaller $d$ the
achievable rate stays constant. This shows us that when the
interferer is close enough, MLE decoding actually performs a kind of
multi-user detection, successfully extracting the interference and
preventing a further drop in performance.

We further compare the achievable rate on a channel with interferer
to the achievable rate without interferer (or equivalently
sufficiently large $d$, e.g. $d=100$m). We see that even when the
interferer is much closer than the transmitter, the rate drop is not
significant. In our examples, it is never more than 50\%. This is in
contrast to \cite{Durisi03} where the rate drops to zero if
interference is much stronger than the signal itself.

\section{Conclusions and Future Work}
\label{sec:concl}

We analyzed the performance of coherent IR-UWB channel with MLE
detector. We presented a novel procedure to calculate a lower-bound
on achievable rates using random-coding techniques and Monte-Carlo
simulations. Using this bound we are able to show that the
performance of an MLE detector is significantly better than the
performance of a widely used nearest-neighbor detector in presence
of a strong impulsive interference. This suggested that the use of a
more complex MLE detector may completely eliminate the need for a
medium access protocol in IR-UWB networks. The analysis of more
complex networking scenarios and different medium access protocols
is left for future work.

\bibliography{main}

\begin{thebibliography}{10}
\providecommand{\url}[1]{#1}
\def\UrlFont{\rmfamily}
\providecommand{\newblock}{\relax}
\providecommand{\bibinfo}[2]{#2}
\providecommand\BIBentrySTDinterwordspacing{\spaceskip=0pt\relax}
\providecommand\BIBentryALTinterwordstretchfactor{4}
\providecommand\BIBentryALTinterwordspacing{\spaceskip=\fontdimen2\font plus
\BIBentryALTinterwordstretchfactor\fontdimen3\font minus
  \fontdimen4\font\relax}
\providecommand\BIBforeignlanguage[2]{{%
\expandafter\ifx\csname l@#1\endcsname\relax
\typeout{** WARNING: IEEEtran.bst: No hyphenation pattern has been}%
\typeout{** loaded for the language `#1'. Using the pattern for}%
\typeout{** the default language instead.}%
\else
\language=\csname l@#1\endcsname
\fi
#2}}

\bibitem{Durisi03}
G.~Durisi and S.~Benedetto, ``Performance evaluation of {TH-PPM UWB} systems in
  the presence of multiusers interference,'' \emph{IEEE Communication Letters},
  vol.~7, no.~5, May 2003.

\bibitem{lapidoth96}
A.~Lapidoth, ``Nearest neighbor decoding for additive non-gausian noise
  channels,'' \emph{IEEE Transactions on Information Theory}, vol.~42, no.~5,
  pp. 1520--1529, September 1996.

\bibitem{Durisi02}
G.~Durisi and G.~Romano, ``On the validity of gaussian approximation to
  characterize the multiuser capacity of {UWB TH-PPM},'' in \emph{Proc. UWBST},
  2002.

\bibitem{Merz05}
R.~Merz and J.-Y. Le~Boudec, ``Conditional bit error rate for an impulse radio
  {UWB} channel with interfering users,'' in \emph{Proc. ICUWB}, 2005.

\bibitem{Souilmi03I}
Y.~Souilmi and R.~Knopp, ``On the achievable rates of ultra-wideband systems in
  multipath fading environments,'' in \emph{ISIT}, July 2003.

\bibitem{Souilmi03D}
Y.~Souilmi and K.~Raymond, ``Challenges in {UWB} signaling for ad-hoc
  networking,'' in \emph{DIMACS Series}, November 2003, pp. 271--284.

\bibitem{Luo06}
X.~Luo and G.~Giannakis, ``Achievable rates of transmitted-reference
  ultra-wideband radio with {PPM},'' \emph{IEEE Transactions on
  Communications}, vol.~54, no.~9, pp. 1536--1541, September 2006.

\bibitem{Steiner05}
C.~Steiner and K.~Witrisal, ``Multiuser interference modeling and suppression
  for a multichannel differential {IR-UWB} system,'' in \emph{ICUWB}, 2005.

\bibitem{Tse05}
D.~Tse and P.~Viswanath, \emph{Fundamentals of Wireless Communication}.\hskip
  1em plus 0.5em minus 0.4em\relax Cambridge University Press, 2005.

\bibitem{Telatar00}
I.~E. Telatar and D.~N.~C. Tse, ``Capacity and mutual information of wideband
  multipath fading channels,'' \emph{IEEE Transactions on Information Theory},
  vol.~46, no.~4, pp. 1384--1400, July 2000.

\bibitem{Win00}
M.~Win and R.~Scholtz, ``Ultra-wide bandwidth time-hopping spread-spectrum
  impulse radio for wireless multiple-access communications,'' \emph{IEEE
  Transactions on Communications}, vol.~48, no.~4, pp. 679--691, April 2000.

\bibitem{LeBoudec06}
\BIBentryALTinterwordspacing
J.-Y. Le~Boudec, ``Performance evaluation,'' EPFL,'' Lecture Notes, 2006.
  [Online]. Available: \url{http://ica1www.epfl.ch/perfeval/}
\BIBentrySTDinterwordspacing

\bibitem{ghassemzadeh03}
S.~Ghassemzadeh and V.~Tarokh, ``Uwb path loss characterization in residential
  environments,'' in \emph{IEEE Radio Frequency Integrated Circuits (RFIC)
  Symposium}, June 2003, pp. 501--504.

\end{thebibliography}
\bibliographystyle{IEEEtran}

\appendix
\label{sec:appendix}

In the appendix we explain how to calculate the channel output
distribution $P(\bR=\bY\,|\,\bV,\bh_1)$ and $J(\bV,\bW,\bh_1) =
\int_\bY P(\bR=\bY\,|\,\bV,\bh_1) P(\bR=\bY\,|\,\bW,\bh_1) d\bY$.
First, conditional to the channel realisation $\bH$ and the
transmitted symbols $\bU$, the channel outputs $R = \{r_m[w]\}$ are
Gaussian i.i.d. RV with distribution
\begin{eqnarray*}
  \lefteqn{P(\bR = \bY|\bU, \bH) = \left({1\over\sqrt{2\pi\sigma_W^2}}\right)^{M N} \times} \\
    &\times& \exp\left(-\sum_{w=1}^{N}\sum_{m=1}^M(y_m[w] - \sum_{i=1}^I v_i[w]A_ih_{im})^2 / 2\sigma_W^2\right).
\end{eqnarray*}
Also, each channel response $\bh_i$ is multivariate Gaussian with distribution
\[
  P(\bH_{-1}) = \left({1\over \sqrt{(2 \pi)^{M} |T|}}\right)^{I-1}
     \exp\left(-{1\over 2}\sum_{i=2}^I \bh_i^T T^{-1} \bh_i\right),
\]
Thus, we have
\begin{eqnarray*}
  \lefteqn{P(\bR = \bY\,|\,\bV, \bh_1) = \EE_{\bH_{-1}} (P(\bR = \bY|\bV, \bH))=}\\
   &=& \int_{\bH_{-1}} \left({1\over \sqrt{(2 \pi)^{M} |T|}}\right)^{I-1}
      \left({1\over\sqrt{2\pi\sigma_W^2}}\right)^{M N} \times\\
   &\times& \exp\left(- \sum_{i=1}^I {\bh_i^T T^{-1} \bh_i \over 2}\right)\times\\
   &\times& \exp\left(- \sum_{w=1}^{N_1} \sum_{m=1}^M{(y_m[w] - \sum_{i=1}^I v_i[w]A_ih_{im})^2 \over 2\sigma_W^2}\right) d\bH_{-1}
\end{eqnarray*}
which is again a multivariate Gaussian and can be expressed in 
closed form.  Similarly, since $P(\bR=\bY\,|\,\bV,\bh_1)$ is
exponential, $J(\bV,\bW,\bh_1)$ is also exponential and can be
calculated explicitly. However, both expressions are long and we do
not give them here.

\end{document}